\documentclass[letterpaper, 10 pt, conference]{ieeeconf}  

\IEEEoverridecommandlockouts                              

\usepackage{xcolor}
\usepackage{soul}
\usepackage{booktabs}
\usepackage{amsmath}
\usepackage{multirow}
\usepackage{titlesec}
\usepackage{graphicx}
\usepackage[linesnumbered,ruled,vlined]{algorithm2e}

\title{\LARGE \bf

A Meaningful Human Control Perspective on User Perception of Partially Automated Driving Systems: A Case Study of Tesla Users

}

\author{Lucas Elbert Suryana$^{1}$, Sina Nordhoff$^{1}$, Simeon C. Calvert$^{1}$,\\ Arkady Zgonnikov$^{2}$, Bart van Arem$^{1}$
\thanks{$^{1}$Lucas E. Suryana, Sina Nordhoff, Simeon C. Calvert, and Bart van Arem are with the Department of Transport and Planning, TU Delft, 2628 CN Delft, The Netherlands. {\tt\small l.e.suryana@tudelft.nl; s.nordhoff@tudelft.nl; s.c.calvert@tudelft.nl; and b.vanarem@tudelft.nl}}%
\thanks{$^{2}$ Arkady Zgonnikov is with the Department of Cognitive Robotics, TU Delft,
        2628 CN Delft, the Netherlands
        {\tt\small a.zgonnikov@tudelft.nl}}%
}

\begin{document}

\maketitle
\thispagestyle{empty}
\pagestyle{empty}

\begin{abstract}

The use of partially automated driving systems raises concerns about potential responsibility issues, posing risk to the system safety, acceptance, and adoption of these technologies. The concept of meaningful human control has emerged in response to the responsibility gap problem, requiring the fulfillment of two conditions, tracking and tracing. While this concept has provided important philosophical and design insights on automated driving systems, there is currently little knowledge on how meaningful human control relates to subjective experiences of actual users of these systems. To address this gap, our study aimed to investigate the alignment between the degree of meaningful human control and drivers’ perceptions of safety and trust in a real-world partially automated driving system. We utilized previously collected data from interviews with Tesla ``Full Self-Driving'' (FSD) Beta users, investigating the alignment between the user perception and how well the system was tracking the users' reasons. We found that tracking of users' reasons for driving tasks (such as safe maneuvers) correlated with perceived safety and trust, albeit with notable exceptions. Surprisingly, failure to track lane changing and braking reasons was not necessarily associated with negative perceptions of safety. However, the failure of the system to track expected maneuvers in dangerous situations always resulted in low trust and perceived lack of safety. Overall, our analyses highlight alignment points but also possible discrepancies between perceived safety and trust on the one hand, and meaningful human control on the other hand. Our results can help the developers of automated driving technology to design systems under meaningful human control and are perceived as safe and trustworthy.

\end{abstract}

\section{INTRODUCTION}\label{introduction}

The increasing use of automated driving systems raises concerns about potential responsibility issues, which can impact safety, adoption, and acceptance of these systems~\cite{nyholm2020automated, calvert2020gaps}. In particular, delegating control to automated systems, either partially or fully, could create responsibility gaps —- situations where no human agent is responsible for the behavior of the system \cite{matthias2004responsibility, santoni2021four}. The concept of meaningful human control (MHC) recently gained prominence in addressing the responsibility gap problem~\cite{santoni2018meaningful, de2022realising, mecacci2023human}. 
This concept posits that humans, not artificial agents, should remain morally responsible for the behavior of automated systems~\cite{santoni2018meaningful}. This entails that (partially or fully) automated systems should be designed in such a way that humans interacting with the systems maintain some form of \textit{meaningful} control over the system behavior, even when not in operational control of the system~\cite{santoni2018meaningful}. Recent work on operationalizing meaningful human control made steps towards specific frameworks and design principles for developing automated driving systems~\cite{heikoop2019human, calvert2020human, cavalcante2023meaningful}. However, there is currently a lack of understanding of how meaningful human control relates to subjective experiences of actual humans, in particular human drivers/users of driving automation.  

In this paper, we aim to investigate the alignment between the degree of meaningful human control and drivers' perceptions of safety and trust in a real-world partially automated driving system. To this end, analyzing subjective evaluation data from participants with real driving experience is crucial. 
In this research, we analyzed previously collected data from interviews with Tesla FSD Beta users~\cite{nordhoff2023drivers1} from the meaningful human control perspective. 
From this data, we gathered information when the participants indicated their perception of trust and safety while describing the behavior of the automated driving technology. Then, we analyzed that information by classifying whether the behavior of the vehicles was \textit{tracking} the users' reasons, along with the corresponding perception of trust and safety. 


\section{Related work}\label{sec:related}
\subsection{Meaningful human control of automated driving systems}
Santoni de Sio \& van den Hoven~\cite{santoni2018meaningful} provide a comprehensive philosophical account of two conditions for a system to be under meaningful human control. First, the \textit{tracking} condition requires that a system should be capable of responding to relevant moral, strategic, and intentional reasons of humans in the environment where the system operates. Second, the \textit{tracing} condition implies that a system should be designed in a way that allows tracing back the outcome of its operation to at least one human in the loop of control. The tracking condition has been operationalized for partially automated driving systems~\cite{mecacci2020meaningful}, connecting strategic and tactical reasons with Michon's classical theory of the driving task \cite{michon1985critical}.  In this approach, the tracking condition is satisfied when the behavior of the automated driving system aligns with the moral (e.g. respecting regulations), strategic (e.g., going home), tactical (e.g., overtaking), and operational (e.g., steering) reasons of relevant humans (e.g., the driver). For instance, if a driver has a tactical reason to change lanes smoothly, the system should perform a lane change considered smooth by the driver. Findings from \cite{mecacci2020meaningful} have then been used as a basis for design guidelines on human-automation interaction in mixed-traffic~\cite{mecacci2023human}. Furthermore, \cite{calvert2020conceptual} applied these findings to create a quantitative formulation for vehicle control. 

\subsection{User perception of safety and trust in automated driving systems}
Studies consistently demonstrate that drivers' perceptions, particularly their perceptions of safety and trust, significantly influence their willingness to adopt automated driving systems \cite{ljubi2023role, montoro2019perceived}. However, such studies have so far been mostly limited to driving simulator experiments and public surveys \cite{koglbauer2018autonomous, bellet2022interaction}, which warrants caution  in assuming that the results accurately reflect real-world driving experiences. A systematic review of driving simulator experiments revealed variations in the representations \cite{wynne2019systematic}. Furthermore, many survey studies primarily relied on drivers' imaginative perceptions and expectations of automated vehicles, lacking practical experience and detailed knowledge about the vehicles' functionality \cite{lee2023does}. Survey studies also have disadvantages in terms of information access, reliability, and validity \cite{burcu2000comparison}. For this reason, in this work we focus on semi-structured interview data collected previously~\cite{nordhoff2023drivers1} that avoids these issues but also provides in-depth insight into the tactical reasons of users.

\section{METHODOLOGY}\label{methodology}

\subsection{Vehicle behavior}
In this work, we connect the concept of meaningful human control to perceived safety and trust via the \textit{behavior} of automated driving systems. According to \cite{nordhoff2024trust}, automation capabilities, including behaviors such as lateral-longitudinal control and collision avoidance, influence the level of perceived safety and trust. These behaviors closely parallel the behaviors of a system that should adhere to the tracking condition, aligning with relevant human reasons. To the best of our knowledge, no existing automated driving systems have been designed with the concept of meaningful human control in mind. Nevertheless, any automated driving system inherently satisfies the tracking condition to some extent \cite{mecacci2020meaningful}. Therefore, our focus in this paper will be on the tracking condition. When we refer to reasons being tracked or not tracked in subsequent sections, it implies whether the vehicle can fulfill the expected tactical or operational reasons of the driver.

The vehicle behavior analyzed in this research is related to the driving tasks of vehicles equipped with automated driving technology. When drivers discussed how the vehicle performed the driving task, it indicated whether the vehicle tracked their reasons or not. Given that the subject of this research is SAE Level 2 vehicles, our focus was on examining the driving tasks that automated driving technologies can perform according to the SAE standard \cite{sae2018taxonomy}. According to this standard, vehicles should have driving assistance features, such as adaptive cruise control and lane centering, at the same time. However, most of the participants did not explicitly mention the names of these features when explaining the behavior of FSD Beta and standard Autopilot and their perceived safety and trust. Therefore, we only highlighted driving tasks that encompass these features, namely steering, braking, and accelerating.

\subsection{Data}
In this study, we utilized transcribed conversation data from \cite{nordhoff2023drivers1}. Our analysis covered aspects that were not discussed in previous research that used this data \cite{nordhoff2023drivers1, nordhoff2023drivers2, nordhoff2024trust}. The data comprised responses from 103 respondents in the FSD Beta program, collected through interviews using a semi-structured protocol that included a total of 35 questions, comprising both open-ended and closed-ended questions. The questions consisted of five sections: general experience, perceptions of vehicle operation, perceptions of safety, exploration of trust level, and typical vehicle usage. The interviews were conducted via Zoom and the participants were recruited through social media.  On average, each interview lasted 78 minutes and yielded approximately 12,200 words. Following quality checks, which involved the removal of non-English interview data and addressing missing transcriptions, only 99 interview data were deemed suitable for further analysis in this research. We only used the answers to the open-ended questions because the purpose was to look for reasons that were only mentioned when the participants explained them. Throughout the interviews, both video and conversation were recorded. 
The interviewer's role was limited to reduce the possibility of bias.

\subsection{Data processing}
Based on the collection of documents where the transcribed conversation data was saved, we further processed the data so that it could be used for further analysis in our keyword search algorithm. We processed each transcript by applying a series of text preprocessing steps: removing newline characters, adding spaces between digits and alphabets, and combining these operations to create a cleaned version of the conversation. The text was then tokenized using the `$word\_tokenize$` function in NLTK (https://www.nltk.org) to break it into individual words. Subsequently, a text cleaning function was applied, removing short words, eliminating spurious characters, transforming letters to lowercase, and removing stopwords and specific words defined in the process, such as the names of participants. The final step involved lemmatization, which normalizes the tokens using NLTK's WordNet lemmatizer. To illustrate, consider a sample text such as "\textit{So changing lanes and avoiding blind spot collisions is very, very good here. I'm maintaining speeds and safe speeds.}". After applying the data processing, the lemmatized tokenized data results will be formatted as a list: [\textit{'changing', 'lane', 'avoiding', 'blind', 'spot', 'collision', 'good', 'maintaining', 'speed', 'safe', 'speed'}].

\subsection{Seed words and keyword search}
To identify instances in the transcribed conversation data where interview participants mentioned the driving tasks and their perceived safety and trust, we employed seed words and a keyword search algorithm with the data. Seed words were defined as individual terms associated with driving tasks, perceived safety, and trust. For driving tasks, we defined the seed words as the verb corresponding to each driving task, except for steering. The choice of the seed word "steer" was often connected to the expression "steering wheel," which was not our focus. Therefore, we sought alternative terms describing the steering process, namely 'change lane' and 'keep lane'. Defining seed words is an iterative step, aligning with \cite{watanabe2022theory}'s suggestion to derive seeds based on the researcher's subject knowledge. For perceived safety and trust, we adapted the seed words used by \cite{nordhoff2024trust}. Subsequently, our keyword search algorithm, which operates by using the defined seed words as keywords, was employed to identify instances in each conversation where participants mentioned these seed words. We chose keyword search over manual inspection because relevant content could potentially span across various answers to different questions, and the considerable length of the conversations made manual inspection impractical. The seed words and the algorithm can be found in Table \ref{table_seed_words} and Algorithm \ref{keyword_search_algorithm}. In Table \ref{table_seed_words}, the use of 'and' indicates that both words must be present for the keyword search algorithm, while the use of 'or' indicates that the presence of either one is sufficient. As an example, assume we have a list of lemmatized tokenized data: \textit{['vehicle', 'good', 'brake', 'really', 'hard', 'try', 'make', 'left', 'turn', 'reason', 'turn', 'blinker', 'quick', 'definitely', 'make', 'feel', 'unsafe', 'house', 'spot', 'signal', 'make', 'better', 'before']}. We are looking for words discussing perceived safety and trust, particularly those related to braking. We set \textit{'brake'} as the seed word for braking and \textit{'safe,' 'happy,' 'relax,' 'comfort,' 'glad,' 'trust,' 'distract,', 'rely,'} and \textit{'trustworthy'} as seed words for perceived safety and trust. The algorithm goes through the list, searching for occurrences of these seed words. When it finds one occurrence in one of the seed words, for instance, the word 'brake', it will expand the search over the next 20 words, resulting in a new list of words \textit{['brake', 'really', 'hard', 'try', 'make', 'left', 'turn', 'reason', 'turn', 'blinker', 'quick', 'definitely', 'make', 'feel', 'unsafe', 'house', 'spot', 'signal', 'make']}. Then, the algorithm checks for any words related to the seeds of perceived safety and trust, and it identifies the word 'safe' inside the term 'unsafe'. Now, there are two occurrences of seed words in the list, and the list will be retrieved. 

After defining seed words and implementing keyword search, we qualitatively classified perceptions by analyzing the word context surrounding the keywords in the transcribed conversations. The classification process began with an initial assessment conducted by the first author. Subsequently, to minimize subjectivity in categorization, discussions were held with the co-authors to ensure alignment with others' perspectives. These discussions involved verifying whether the systems tracked participants' reasons during actions such as braking and determining whether the braking process indicated positive or negative perceptions of safety and trust. 

\begin{table}[]
\caption{Seed words}
\label{table_seed_words}
\begin{tabular}{|l|l|l|}
\hline
\textbf{Category}             & \textbf{Sub-category} & \textbf{Seed words}                                           \\ \hline
\multirow{4}{*}{Driving taks} & Accelerating          & 'accelerate'                                                  \\ \cline{2-3} 
& Braking               & 'brake'                                                       \\ \cline{2-3} 
& Lane changing         & 'lane' and 'change'                                           \\ \cline{2-3} 
& Lane keeping          & 'lane' and 'keep'                                             \\ \hline
Perceived safety              & -                     & \begin{tabular}{@{}c@{}} 'safe' or 'happy' or 'relax' \\ or 'comfort' or 'glad'\end{tabular}           \\ \hline
Trust                         & -                     & \begin{tabular}{@{}c@{}} 'trust' or 'distract' or 'rely' \\ or 'trustworthy' or 'comfort'\end{tabular} \\ \hline
\end{tabular}
\end{table}



\begin{algorithm}
    \SetAlgoNlRelativeSize{0}
    \KwData{Seed Words ($\text{seeds}$), Tokenized Data ($\text{tokenList}$), Buffer Size ($\text{bufferSize}$)}
    \KwResult{Retrieved list ($\text{list}$)}

    $\text{bufferSize} \gets 20$\;
    $\text{list} \gets []$\;
    $\text{threshold} \gets \sum ($\text{seed} \text{in} \text{seeds}$)$\;

    \For{$\text{token}, \text{index}$ \textbf{in} $\text{tokenList}$}{
        \If{\text{token} \textbf{in} \text{seeds}}{
            $\text{tokenBuffer} \gets \text{tokenList}[index:index+\text{bufferSize}]$\;

            $\text{seedCount} \gets \sum_{\text{seed} \text{ in } \text{seeds}} (\text{seed} \text{ in } \text{tokenBuffer})$\;

            \If{\text{seedCount} $>$ \text{threshold}}{
                $\text{list} \gets \text{tokenBuffer}$\;
            }
        }
    }
    \caption{Keyword Search Algorithm}
    \label{keyword_search_algorithm}
\end{algorithm}

\section{RESULTS}\label{results}

After performing the keyword search algorithm and qualitative classification, we obtained the results displayed in Table \ref{table_classification_result}. The first column represents the driving tasks of the automated driving systems. The second column classifies perceived safety and trust into four categories: safe, unsafe, trust, and lack of trust. The numbers in the third and fourth columns indicate whether the vehicle satisfied the tracking condition of meaningful human control (i.e., tracked or did not track the participant's reasons).

In addition to summary statistics (Table \ref{table_classification_result}), we qualitatively analyzed the instances of participants' perceptions of driving tasks performed by automated driving systems, along with details regarding whether their reasons were tracked or not. As the primary focus of this research was to investigate the alignment between tracking and perceived safety and trust, we specifically chose parts of the results that fall within the same category for each perception for further analysis. We developed three categories by examining the pattern of reasons tracked and not tracked for each perception. The first category, 'Controversy,' includes perceptions where both reasons were tracked and not tracked. The second and third categories, 'Reasons mostly tracked' and 'Reasons mostly not tracked,' show perceptions where the majority of reasons were either tracked or not tracked.

\begin{table}[]
\centering
\caption{Driving task and user perception analysis results}
\label{table_classification_result}
\begin{tabular}{|l|l|l|l|}
\hline
\textbf{Driving tasks} & \textbf{Perception} & \textbf{\begin{tabular}[c]{@{}l@{}}Reasons \\ tracked\end{tabular}} & \textbf{\begin{tabular}[c]{@{}l@{}}Reasons  \\ not tracked\end{tabular}} \\ \hline
\multirow{4}{*}{Accelerating}                                            & Safe                & 2                                                                   & 2                                                                        \\ \cline{2-4} 
& Unsafe              & 0                                                                   & 8                                                                        \\ \cline{2-4} 
& Trust          & 1                                                                   & 0                                                                        \\ \cline{2-4} 
& Lack of trust          & 0                                                                   & 4                                                                        \\ \hline
\multirow{4}{*}{Braking}                                                 & Safe                & 19                                                                  & 7                                                                        \\ \cline{2-4} 
& Unsafe              & 0                                                                   & 21                                                                       \\ \cline{2-4} 
& Trust          & 8                                                                   & 2                                                                        \\ \cline{2-4} 
& Lack of trust          & 1                                                                   & 5                                                                        \\ \hline
\multirow{4}{*}{\begin{tabular}[c]{@{}l@{}}Lane\\ changing\end{tabular}} & Safe                & 9                                                                   & 4                                                                        \\ \cline{2-4} 
& Unsafe              & 0                                                                   & 5                                                                        \\ \cline{2-4} 
& Trust          & 4                                                                   & 0                                                                        \\ \cline{2-4} 
& Lack of trust          & 0                                                                   & 0                                                                        \\ \hline
\multirow{4}{*}{\begin{tabular}[c]{@{}l@{}}Lane \\ keeping\end{tabular}} & Safe                & 11                                                                  & 0                                                                        \\ \cline{2-4} 
& Unsafe              & 0                                                                   & 3                                                                        \\ \cline{2-4} 
& Trust          & 9                                                                   & 0                                                                        \\ \cline{2-4} 
& Lack of trust          & 0                                                                   & 0                                                                        \\ \hline
\end{tabular}
\end{table}

\subsection{System perceived as safe}
The results showed that when the participants perceived the driving tasks of accelerating, braking, and lane changing as safe, there were instances where their reasons were tracked and not tracked. 

\subsubsection{Controversy: lane changing}
The participants mentioned thirteen instances where they felt safe with the systems. Nine instances indicated that the systems tracked participants' reasons, while the rest did not. One participant emphasized that significant improvements in automatic lane changes contributed to their sense of safety. Previously, their vehicle would perform several unsafe lane changes, but now it has become as safe as their own driving. Another participant simply did not want to be bothered with lane-changing, as they believed that the systems worked properly and safely.

\begin{itemize}
    \item \textit{"I do feel safe when it's doing automatic lane changes, and that's a massive improvement since when it first was released, it used like almost every autopilot lane change when it first came out. Would be cutting someone off. Pulling into a lane with someone rapidly approaching or some other form of unsafe lane change now..The automatic lane change feature is.. about as safe as I am. (R007)"}
    \item \textit{"I don't want to be bothered with changing lanes.. So from my perspective, let the car do it. I know the car is safe, it will do it properly. (R035)"}
\end{itemize}

Four participants mentioned that they felt safe even though their reasons were not tracked. One of them described how the vehicles kept their turn signals differently from human drivers, but they believed it was because the system was a better driver. Another participant felt safe and shared an occasion where they were impressed with how the system took a lane. They were initially unsure if it was a faster lane, but it turned out to be correct several seconds later.

\begin{itemize}
    \item \textit{"I feel safe when I am on the freeway.. it's s better driver than I am at taking turns on the freeway..  Changing lanes.. I like how it keeps his turn signal on all the way through the entire lane change and sometimes human drivers will just do a couple blinks. (R043)"}
    \item \textit{"I do feel safe with autopilot.. but I feel like it's more of a clever thing like it makes really interesting decisions.. like changing into a lane that doesn't yet appear to be faster, but it says it's changing into a faster lane and then 10 seconds later it is the faster lane.. autopilot picked the correct lane and it does it so many times it's like it can't be coincidence. (R074)"}
\end{itemize}

\subsubsection{Controversy: braking}
Participants mentioned nineteen instances where the system tracked their reasons. One participant emphasized feeling safe due to the automated driving system's ability to slam the brakes faster than they could react. Another participant indicated a feeling of relaxation, which is linked to the perception of safety, as it reduced repetitive driving tasks during traffic jams. For example:

\begin{itemize}
    \item \textit{"Completely safe.. You know, it slammed on the brakes when the guy in front of me slammed on the brakes, faster than I could. (R084)"}
    \item \textit{"It can just kind of help you relax a bit.. one time I was stuck in a traffic jam and autopilot was nice because.. I didn't have to.. put the gas on and then put the brake on and then put the gas on in there.. like over and over and over again. (R059)"}
\end{itemize}

However, seven instances were perceived as safe situations even when participant's reasons were not tracked. Two participants felt safe in braking scenarios, like sudden brakes or brakes that were too slow in their perception, because they could quickly take over. For example:

\begin{itemize}
    \item \textit{"I keep my foot almost always on the gas pedal so that if it brakes suddenly, I can quickly override it.. That's how I feel very safe. (R047)"}
    \item \textit{"So if I'm coming up on a stop sign and it's not slowing down soon enough for what I would expect, then I will manually hit the brakes and disengage.. I've never felt unsafe though and using it. (R087)"}
\end{itemize}

\subsubsection{Reasons mostly tracked: lane keeping}
When the participants perceived lane keeping as safe, all instances showed that their reasons were tracked. Participants mentioned eleven instances where the automated driving system's lane keeping tracked their reasons and gave them safe perceptions. One participant described that they felt safe because the systems reduced their main workload so that they did not have to pay attention to lane keeping. Another participant mentioned that the systems did a very good job of lane keeping. For example:

\begin{itemize}
    \item \textit{"I feel like Autopilot makes me safer because it reduces my workload.. in terms of lane keeping mainly.. It makes me feel more comfortable as a driver just because I don't have to pay attention to the lines on the road. (R024)"}
    \item \textit{"I feel safe when autopilot and FSD beta..they're not perfect, but yes. They do a very good job of keeping you in your lane. (R099)"}
\end{itemize}

\subsection{System perceived as unsafe}
When the participants perceived all driving tasks as unsafe, all instances showed that their reasons were not tracked.

\subsubsection{Reasons mostly not tracked: lane changing}
When the participants explained the situations in which they felt unsafe while changing lanes, the automated driving systems did not track their reasons at all. One of the participants mentioned that they did not feel safe because the system did not have human-like behaviors. They further emphasized that it was too fast or too close to other vehicles at stop signs or traffic lights. Another participant highlighted that they could feel safer if the systems could adapt to how humans drive and incorporate a more human-like feeling into them.

\begin{itemize}
    \item \textit{"How do you feel when you feel safe or unsafe?.. once they start getting to the point where it has more human like behaviors and doesn't have those.. conflicts in what it's seeing and hitting the brakes or making a turn, you know, changing lanes. (R081)"}
    \item \textit{"Having a limit on acceleration when changing lanes.. might make sense in kind of a textbook way of safety driving.. it does not match up with the way that humans drive and how you should be adapting to how humans drive.. putting in some more.. natural feeling or more human feeling.. would make me feel safer." (R050)}
\end{itemize}

\subsubsection{Reasons mostly not tracked: accelerating}
No reasons were tracked when the participants indicated that they felt unsafe with the automated driving system's acceleration. Out of eight instances where they described feeling unsafe, their reasons were not tracked. One of them mentioned that the vehicle did not feel like a safe driver because it did not blend its speed to accelerate with the flow of traffic. Another participant felt unsafe when merging on the highway because the vehicle did not do a good job of accelerating to match the other vehicle.

\begin{itemize}
    \item \textit{"It would be safer for the car to accelerate more just with the flow of traffic.. just being able to kind of blend in with the drivers around you, I think as part of being a safe driver. And so since our still situations where the car will not speed up when appropriate or when safe, I'm going to say that there are safety issues. (R050)"}
    \item \textit{"But when I'm merging on to the highway. That's when I feel the most unsafe because it it will get on to the on ramp. Accelerate to match the speed of the other vehicles, which it doesn't really good job of. (R074)"}
\end{itemize}

\subsection{Trust}
Our results indicate that participants had trust in the automated driving system's braking in eight instances when their reasons were tracked and in two instances when their reasons were not tracked.

\subsubsection{Controversy: braking}
The participants described instances when they had high and low levels of trust in the braking experience. When they mentioned that they had trust in the automated driving systems, their reasons were tracked eight times, and two times their reasons were not tracked. Things that made the participants trust the automated driving systems were their capability to brake according to the traffic rules and brake to stop better than themselves.

\begin{itemize}
    \item \textit{"One advantage of the larger Autopilot is that it can automatically stop at traffic lights. That works pretty well too. It's comfortable. (R047)"}
    \item \textit{"Stays safe distances I have been in on-air state whenever they just slammed on their brakes and it has stopped very well. Probably better and I would have done. I would have panic stopped and it stopped perfectly. Maintains a good distance. Ohh. So yes, I trust it. (R062)"}
\end{itemize}

However, there were also instances when they had trust even though the vehicle did not track their reasons. One participant mentioned that even though they intervened in some instances where they felt unsafe, they felt more comfortable the more they drove because the level of perceived safety changed over time. Another participant described a situation where they hit the brake to avoid a long truck but still felt comfortable with the system. They attributed this comfort to the benefit of feeling less tired, even though they still needed to monitor the system.

\begin{itemize}
    \item \textit{"So whenever I feel unsafe or even uncertain, I'll.. tap the brakes or move the stock up.. it's not difficult for me to do that.. I go in and out of FSD all the time, and that's how I handle this issue of when I don't feel safe.. As your perceived safety changed over time.. the more I drive it, the more comfortable I get." (R103)}
    \item \textit{"That truck tried to turn into my lane and the car. I didn't realize that it was a really long trailer. So I had to just slow it down, hit the brake, but so you can get pretty comfortable with it and pretty relaxed. I would say you know, so still watching, but it's it actually you're less tired when you get where you're going." (R100)}
\end{itemize}

\subsubsection{Reasons mostly tracked: lane changing}
Nevertheless, when the participants had trust in the driving tasks of accelerating, lane changing, and lane keeping, all instances indicated that all of the participants' reasons were tracked. During lane-changing situations, participants mentioned four instances where their reasons were tracked and they had a higher level of trust. One participant initially faced trust issues with the system because it frequently placed them in the wrong lane. However, after an update that required confirmation before lane changes and smooth experiences, the participant's trust started to build. Another participant described a trust-building process with a software upgrade, allowing the system to perform complex maneuvers in city traffic.

\begin{itemize}
    \item \textit{"You frequently get in the wrong lane.. And so I did all the driving through there. Uh, because I didn't trust Autopilot.. but by then I gained some confidence in the lane changing.. so I let it tell me when to change lanes and I just confirmed it.. do the lane changing. It was the smoothest trip I've ever taken. (R010)"}
    \item \textit{"I got the full self-driving upgrade software.. and then I had to get used to it. Changing lanes had to get used to it, doing on ramps and off ramps.. And then I went to FSD beta. And then now I've got to be able to.. trust and get used to the car doing city traffic and these very complex interceptions, these complex intervals (R048)"}
\end{itemize}

\subsection{Lack of trust}
When participants expressed lower trust in the automated driving system's braking, one instance showed that their reasons were tracked, while five instances indicated that their reasons were not tracked.

\subsubsection{Reasons mostly not tracked: braking}
Participants who felt lack of trust described that the systems failed to track their reasons. They mentioned that they did not trust the automated driving systems because the systems could make unexpected wrong decisions very fast, such as braking when there was a pedestrian on the crosswalks. Another participant described that they did not trust the systems when entering a highway because the systems did not brake when there was not much of a gap.

\begin{itemize}
    \item \textit{"I don't fully trust it when it's active.. it can just do something wrong very fast.. you just do not expect it. For example, if it sees a pedestrian.. on the crosswalk, it'll kind of just slam on the brakes. (R065)"}
    \item \textit{"You're gonna turn onto that highway.. It will creep forward.. it'll sometimes creep into the other highway. So you have to hit the brakes, or it'll appear to start going when there's not much of a gap, so you slam on the brakes again. So I don't trust it. (R010)"}
\end{itemize}

However, in one instance, the systems tracked their reasons. The only participant who felt lack of trust, even though the systems tracked their reasons, mentioned they admitted Autopilot may react and brake faster than a human, but they felt uncertain because they still needed to rely on themselves.

\begin{itemize}
    \item \textit{"I think the Autopilot may even be better than a human, and if we take emergency braking, then it can probably react and brake faster than me or certainly close to that, but.. you have to assess the situation, then it's uncertain, and that's why I always have to rely on myself. (R015)"}
\end{itemize}

\section{DISCUSSION}\label{discussions}

\subsection{Alignment between tracking and perception of safety}
\subsubsection{Perceived safety}
Our results revealed that participants could feel that lane-changing and braking behaviors of the vehicle were safe, even when the reasons were not being tracked (the `controversy` category). Additionally, we also found the 'reasons mostly tracked' category, where participants perceived lane-keeping tasks as safe, and all their reasons were tracked. Participants who felt safe with lane changing observed better vehicle performance in tracking their reasons, noting that actions like acceleration, deceleration, and lane changes were as safe as manual driving. This aligns with \cite{koglbauer2018autonomous}, where positive experiences with vehicle adaptation led to increased perceived safety and trust. The system's performance in lane-changing, braking, and lane-keeping tasks influenced participants' safety perceptions by meeting expectations and providing a relaxed driving experience with less difficulty and workload. According to \cite{xu2018drives}, reliable experiences with self-driving vehicles increased trust, perceived usefulness, and perceived ease of use. This likely explains why participants believed the system was safe.

On the other hand, feeling safe doesn't always mean that the automated driving systems tracked all their reasons; there were instances of lane-changing and braking tasks where tracking failed. One participant noted that braking was too slow for them, indicating their expected deceleration reason was not tracked. However, they felt safe as long as they were ready to take over. When the driver realized operational reasons for deceleration were not fulfilled, they often took control. The participant also reported no difficulties with the takeover process. \cite{ma2021drivers} found that aggressive drivers were more likely to take control when driving defensively programmed automated vehicles (AVs), potentially explaining the frequent initiation of takeover processes by drivers. The study also revealed that perceived safety levels were similar when aggressive drivers operated both aggressive and defensive AVs. Despite this, the safety score remained higher than when defensive drivers operated aggressive AVs, potentially explaining the continued feeling of safety.

The other instances indicated that the vehicle acted differently from what they expected, but it still led to perceived safety because they believed it was a better driver. The experience of many unexpected maneuvers that turned out to be correct decisions also influenced their belief. The more the vehicles behaved according to the expected reasons, the more they gained trust from the driver. This trust would make people perceive the system as safe, even if it fails to track their reasons. This is in line with the findings on the effect of reliable experiences with trust that the system is safe~\cite{xu2018drives}. Thus, we found indications that tracking driver expectations, such as performing safe maneuvers like lane-keeping, braking, and lane-changing like human drivers, positively correlated with perceived safety. However, failure to track reasons for lane changing and braking was not necessarily associated with perceived lack of safety; it depended on other factors like the numerous reliable experiences, the driver's trust level, and ease of taking over control.

\subsubsection{Perceived lack of safety}
We observed instances where participants felt unsafe during accelerating and lane-changing experiences. Interestingly, in all these instances, the systems failed to track their reasons. Participants noted that the vehicles did not drive in a manner consistent with human behavior. According to \cite{peng2022drivers}, human drivers express a higher comfort level with systems that mimic human driving. In this research, we linked the term 'comfort' to the perception of safety and trust, aligning with our observations. Another instance occurred when the vehicle failed to adjust its speed to match the surrounding vehicles during merging or while on the road. It seems that when the vehicle does not track the driver's reasons in situations involving other vehicles that might lead to dangerous situations, it contributes to the perception of being unsafe. This observation aligns with findings from \cite{borowsky2010age}, which suggest that drivers tend to intensely pay attention to potential dangers with other vehicles in specific traffic situations, such as merging roads. Our findings suggest that the failure of automated driving systems to track drivers' reasons for human-like acceleration and lane-changing behaviors in potentially dangerous traffic situations can contribute to perceived lack of safety.

\subsection{Alignment between tracking and level of trust}
\subsubsection{Trust}
Our findings highlighted that participants could exhibit trust in vehicle's behavior during braking tasks regardless of whether their reasons were tracked (the `controversy` category). Additionally, we observed another category: 'reasons mostly tracked,' where participants expressed higher trust in lane changing, and all of their reasons were tracked. We further examined situations where participants demonstrated a higher level of trust in both braking and lane changing, analyzing scenarios where their reasons were tracked. In these instances, participants emphasized that their positive experiences with the system's performance in intended situations contributed to a higher level of trust. Again, this is in line with the finding from \cite{koglbauer2018autonomous}. Furthermore, the transparency experienced during lane changes, where the driver could anticipate the direction of the vehicle's lane change and had the chance to confirm it, significantly contributed to building the driver's trust. This aligns with a study by \cite{nordhoff2024trust} that indicates transparency positively impacts trust levels. Additionally, it supports findings from \cite{detjen2021towards}, suggesting that using displays to indicate maneuver intentions increases overall transparency in the driving experience.

In contrast, there were instances where participants expressed trust even though the automated driving systems failed to track their reasons for braking tasks. Despite encountering situations where the vehicle made them feel unsafe, the participants maintained their trust in the system due to the ease of taking over and the relaxed experience it provided. When the driver took over control of the system, we expected they did that to reduce the perceived risk they faced. As they had ease of taking over, they could reduce the perceived risk. According to \cite{zhang2019roles} and \cite{he2022modelling}, perceived safety risk has a negative correlation with trust. Thus, these findings might justify why the driver kept feeling safe because they could take over control to reduce the perceived risk, thereby increasing their trust level. Furthermore, the relaxed experience with the system suggests that participants have previously had positive experiences, contributing to a higher level of trust \cite{xu2018drives}. These situations could recover the trust that might temporarily decline during the takeovers or failure of the systems to track the participant's reasons \cite{kraus2020more}.

We found indications that tracking driver expectations, including improved maneuver execution and human-like performance in intended traffic situations in lane-changing and braking tasks, positively relates to trust. Additionally, transparency, which does not relate to tactical or operational reasons, was also positively associated with trust. Notably, the failure to track reasons for braking tasks did not always result in the lack of trust; factors such as positive experiences and the ease of taking over control played a crucial role here.

\subsubsection{Lack of trust}
Our analysis showed situations where participants reported a lower level of trust in braking tasks; their reasons were always not tracked in such cases. Participants expressed concerns that the system could make unexpected decisions, potentially leading to collisions with other vehicles, thereby reducing their trust. In one contrasting case, a participant had lack of trust, but their reasons were tracked. They believed the system could outperform humans in reaction time and braking. However, the uncertainty introduced by the need to assess situations led them to consistently rely on themselves. Manufacturers of automated driving systems explicitly instructed FSD Beta program users to maintain constant attention and be ready to act at any time \cite{nordhoff2023drivers2}. This requirement might explain their lack of trust, as it implies reliance on personal judgment.

\subsection{Limitations}

First, the data in this research focused on drivers' subjective perceptions which may not accurately reflect the actual on-road situation (e.g., findings from \cite{von2020crash} indicate that subjective risk can significantly differ from actual crash risk). Future research should investigate telemetry data in addition to subjective reports. Second, we limited the tactical and operational reasons for relevant human agents only to the driver's perspective. However, many other agents (e.g., vehicle manufacturers and other road users) could and should have meaningful human control over automated driving systems~\cite{mecacci2020meaningful}. We recommend further research to evaluate the tracking of the reasons of other relevant human agents for a more holistic analysis. Third, our results (e.g., Table \ref{table_classification_result}) might not have fully captured the entirety of the participants' responses. The use of seed words may restrict instances that are essentially the same but articulated with different words. We recommend incorporating a more extensive set of alternative seed words, especially in the context of driving tasks, for more comprehensive results. Fourth, our findings are limited by the small number of participants in the FSD Beta program. Future research should aim for a more diverse sample encompassing a wider range of manufacturers and participants to better represent the population. Finally, the qualitative nature of our evaluation presents challenges in scalability, particularly when confronted with a larger database of interviews. Given the expansive volume of data, our current methods may prove impractical to implement.

\section{CONCLUSIONS}\label{conclusions}

This study investigated the alignment between tracking component of meaningful human control and user perception of safety and trust. Successfully tracking drivers' reasons for driving tasks, such as safe maneuvers, performance improvement, and transparency, positively influenced the perception of safety and trust levels. However, the failure to track reasons for lane changing and braking tasks did not necessarily result in negative perception of safety and low trust. Factors such as the ease of taking over control and reliable experiences contributed to drivers feeling safe and maintaining trust. Nevertheless, the failure to track drivers' reasons for expected movements and human-like behaviors in potentially dangerous traffic situations was associated with perceived lack of safety and low trust. Our results can help the developers of automated driving technology to design systems that are under meaningful human control and are perceived as safe and trustworthy.

\section*{Acknowledgments}
We declare that OpenAI ChatGPT was used in the writing process to enhance the style and conciseness of the text originally written by the authors. 

This work was funded by the Indonesia Endowment Fund for Education (LPDP) under Grant 0006552/TRA/D/19/lpdp2021. 








\begin{thebibliography}{10}
\providecommand{\url}[1]{#1}
\csname url@samestyle\endcsname
\providecommand{\newblock}{\relax}
\providecommand{\bibinfo}[2]{#2}
\providecommand{\BIBentrySTDinterwordspacing}{\spaceskip=0pt\relax}
\providecommand{\BIBentryALTinterwordstretchfactor}{4}
\providecommand{\BIBentryALTinterwordspacing}{\spaceskip=\fontdimen2\font plus
\BIBentryALTinterwordstretchfactor\fontdimen3\font minus \fontdimen4\font\relax}
\providecommand{\BIBforeignlanguage}[2]{{%
\expandafter\ifx\csname l@#1\endcsname\relax
\typeout{** WARNING: IEEEtran.bst: No hyphenation pattern has been}%
\typeout{** loaded for the language `#1'. Using the pattern for}%
\typeout{** the default language instead.}%
\else
\language=\csname l@#1\endcsname
\fi
#2}}
\providecommand{\BIBdecl}{\relax}
\BIBdecl

\bibitem{nyholm2020automated}
S.~Nyholm and J.~Smids, ``Automated cars meet human drivers: responsible human-robot coordination and the ethics of mixed traffic,'' \emph{Ethics and Information Technology}, vol.~22, pp. 335--344, 2020.

\bibitem{calvert2020gaps}
S.~C. Calvert, B.~van Arem, D.~D. Heikoop, M.~Hagenzieker, G.~Mecacci, and F.~S. de~Sio, ``Gaps in the control of automated vehicles on roads,'' \emph{IEEE intelligent transportation systems magazine}, vol.~13, no.~4, pp. 146--153, 2020.

\bibitem{matthias2004responsibility}
A.~Matthias, ``The responsibility gap: Ascribing responsibility for the actions of learning automata,'' \emph{Ethics and information technology}, vol.~6, pp. 175--183, 2004.

\bibitem{santoni2021four}
F.~Santoni~de Sio and G.~Mecacci, ``Four responsibility gaps with artificial intelligence: Why they matter and how to address them,'' \emph{Philosophy \& Technology}, vol.~34, pp. 1057--1084, 2021.

\bibitem{santoni2018meaningful}
F.~Santoni~de Sio and J.~Van~den Hoven, ``Meaningful human control over autonomous systems: A philosophical account,'' \emph{Frontiers in Robotics and AI}, vol.~5, p.~15, 2018.

\bibitem{de2022realising}
F.~S. de~Sio, G.~Mecacci, S.~Calvert, D.~Heikoop, M.~Hagenzieker, and B.~van Arem, ``Realising meaningful human control over automated driving systems: a multidisciplinary approach,'' \emph{Minds and machines}, pp. 1--25, 2022.

\bibitem{mecacci2023human}
G.~Mecacci, S.~C. Calvert, and F.~Santoni~de Sio, ``Human--machine coordination in mixed traffic as a problem of meaningful human control,'' \emph{AI \& society}, pp. 1--16, 2023.

\bibitem{heikoop2019human}
D.~D. Heikoop, M.~Hagenzieker, G.~Mecacci, S.~Calvert, F.~Santoni De~Sio, and B.~van Arem, ``Human behaviour with automated driving systems: a quantitative framework for meaningful human control,'' \emph{Theoretical issues in ergonomics science}, vol.~20, no.~6, pp. 711--730, 2019.

\bibitem{calvert2020human}
S.~C. Calvert, D.~D. Heikoop, G.~Mecacci, and B.~Van~Arem, ``A human centric framework for the analysis of automated driving systems based on meaningful human control,'' \emph{TheoreTical issues in ergonomics science}, vol.~21, no.~4, pp. 478--506, 2020.

\bibitem{cavalcante2023meaningful}
L.~Cavalcante~Siebert, M.~L. Lupetti, E.~Aizenberg, N.~Beckers, A.~Zgonnikov, H.~Veluwenkamp, D.~Abbink, E.~Giaccardi, G.-J. Houben, C.~M. Jonker \emph{et~al.}, ``Meaningful human control: actionable properties for ai system development,'' \emph{AI and Ethics}, vol.~3, no.~1, pp. 241--255, 2023.

\bibitem{nordhoff2023drivers1}
S.~Nordhoff and J.~De~Winter, ``Why do drivers and automation disengage the automation? results from a study among tesla users,'' \emph{arXiv preprint arXiv:2309.10440}, 2023.

\bibitem{mecacci2020meaningful}
G.~Mecacci and F.~Santoni~de Sio, ``Meaningful human control as reason-responsiveness: the case of dual-mode vehicles,'' \emph{Ethics and Information Technology}, vol.~22, pp. 103--115, 2020.

\bibitem{michon1985critical}
J.~A. Michon, ``A critical view of driver behavior models: what do we know, what should we do?'' in \emph{Human behavior and traffic safety}.\hskip 1em plus 0.5em minus 0.4em\relax Springer, 1985, pp. 485--524.

\bibitem{calvert2020conceptual}
S.~C. Calvert and G.~Mecacci, ``A conceptual control system description of cooperative and automated driving in mixed urban traffic with meaningful human control for design and evaluation,'' \emph{IEEE Open Journal of Intelligent Transportation Systems}, vol.~1, pp. 147--158, 2020.

\bibitem{ljubi2023role}
K.~Ljubi and A.~Groznik, ``Role played by social factors and privacy concerns in autonomous vehicle adoption,'' \emph{Transport policy}, vol. 132, pp. 1--15, 2023.

\bibitem{montoro2019perceived}
L.~Montoro, S.~A. Useche, F.~Alonso, I.~Lijarcio, P.~Bos{\'o}-Segu{\'\i}, and A.~Mart{\'\i}-Belda, ``Perceived safety and attributed value as predictors of the intention to use autonomous vehicles: A national study with spanish drivers,'' \emph{Safety Science}, vol. 120, pp. 865--876, 2019.

\bibitem{koglbauer2018autonomous}
I.~Koglbauer, J.~Holzinger, A.~Eichberger, and C.~Lex, ``Autonomous emergency braking systems adapted to snowy road conditions improve drivers' perceived safety and trust,'' \emph{Traffic injury prevention}, vol.~19, no.~3, pp. 332--337, 2018.

\bibitem{bellet2022interaction}
T.~Bellet, S.~Laurent, J.-C. Bornard, I.~Hoang, and B.~Richard, ``Interaction between pedestrians and automated vehicles: Perceived safety of yielding behaviors and benefits of an external human--machine interface for elderly people,'' \emph{Frontiers in psychology}, vol.~13, p. 1021656, 2022.

\bibitem{wynne2019systematic}
R.~A. Wynne, V.~Beanland, and P.~M. Salmon, ``Systematic review of driving simulator validation studies,'' \emph{Safety science}, vol. 117, pp. 138--151, 2019.

\bibitem{lee2023does}
Y.~M. Lee, R.~Madigan, T.~Louw, E.~Lehtonen, and N.~Merat, ``Does users’ experience and evaluation of level 3 automated driving functions predict willingness to use: Results from an on-road study,'' \emph{Transportation research part F: traffic psychology and behaviour}, vol.~99, pp. 473--484, 2023.

\bibitem{burcu2000comparison}
A.~Burcu, ``A comparison of two data collecting methods: interviews and questionnaires,'' \emph{Hacettepe Univ J Educ}, vol.~18, pp. 1--10, 2000.

\bibitem{nordhoff2024trust}
S.~Nordhoff and M.~Hagenzieker, ``I will raise my hand and say 'i over-trust autopilot'. i use it too liberally" - driver's reflections on their use of partial driving automation, trust, and perceived safety,'' \emph{Manuscript under review}, 2024.

\bibitem{sae2018taxonomy}
S.~International, ``Taxonomy and definitions for terms related to driving automation systems for on-road motor vehicles,'' \emph{SAE Int.}, vol. 4970, no. 724, pp. 1--5, 2018.

\bibitem{nordhoff2023drivers2}
S.~Nordhoff, J.~D. Lee, S.~C. Calvert, S.~Berge, M.~Hagenzieker, and R.~Happee, ``(mis-)use of standard autopilot and full self-driving (fsd) beta: Results from interviews with users of tesla's fsd beta,'' \emph{Frontiers in Psychology}, vol.~14, 2023.

\bibitem{watanabe2022theory}
K.~Watanabe and Y.~Zhou, ``Theory-driven analysis of large corpora: Semisupervised topic classification of the un speeches,'' \emph{Social Science Computer Review}, vol.~40, no.~2, pp. 346--366, 2022.

\bibitem{xu2018drives}
Z.~Xu, K.~Zhang, H.~Min, Z.~Wang, X.~Zhao, and P.~Liu, ``What drives people to accept automated vehicles? findings from a field experiment,'' \emph{Transportation research part C: emerging technologies}, vol.~95, pp. 320--334, 2018.

\bibitem{ma2021drivers}
Z.~Ma and Y.~Zhang, ``Drivers trust, acceptance, and takeover behaviors in fully automated vehicles: Effects of automated driving styles and driver’s driving styles,'' \emph{Accident Analysis \& Prevention}, vol. 159, p. 106238, 2021.

\bibitem{peng2022drivers}
C.~Peng, N.~Merat, R.~Romano, F.~Hajiseyedjavadi, E.~Paschalidis, C.~Wei, V.~Radhakrishnan, A.~Solernou, D.~Forster, and E.~Boer, ``Drivers’ evaluation of different automated driving styles: Is it both comfortable and natural?'' \emph{Human factors}, p. 00187208221113448, 2022.

\bibitem{borowsky2010age}
A.~Borowsky, D.~Shinar, and T.~Oron-Gilad, ``Age, skill, and hazard perception in driving,'' \emph{Accident analysis \& prevention}, vol.~42, no.~4, pp. 1240--1249, 2010.

\bibitem{detjen2021towards}
H.~Detjen, M.~Salini, J.~Kronenberger, S.~Geisler, and S.~Schneegass, ``Towards transparent behavior of automated vehicles: Design and evaluation of hud concepts to support system predictability through motion intent communication,'' in \emph{Proceedings of the 23rd International Conference on Mobile Human-Computer Interaction}, 2021, pp. 1--12.

\bibitem{zhang2019roles}
T.~Zhang, D.~Tao, X.~Qu, X.~Zhang, R.~Lin, and W.~Zhang, ``The roles of initial trust and perceived risk in public’s acceptance of automated vehicles,'' \emph{Transportation research part C: emerging technologies}, vol.~98, pp. 207--220, 2019.

\bibitem{he2022modelling}
X.~He, J.~Stapel, M.~Wang, and R.~Happee, ``Modelling perceived risk and trust in driving automation reacting to merging and braking vehicles,'' \emph{Transportation research part F: traffic psychology and behaviour}, vol.~86, pp. 178--195, 2022.

\bibitem{kraus2020more}
J.~Kraus, D.~Scholz, D.~Stiegemeier, and M.~Baumann, ``The more you know: trust dynamics and calibration in highly automated driving and the effects of take-overs, system malfunction, and system transparency,'' \emph{Human factors}, vol.~62, no.~5, pp. 718--736, 2020.

\bibitem{von2020crash}
R.~von St{\"u}lpnagel and J.~Lucas, ``Crash risk and subjective risk perception during urban cycling: Evidence for congruent and incongruent sources,'' \emph{Accident Analysis \& Prevention}, vol. 142, p. 105584, 2020.

\end{thebibliography}
\end{document}